\documentstyle[12pt]{article}

\newcommand{\sect}[1]{\section{#1}\setcounter{equation}{0}}

\begin{document}
\begin{titlepage}
\begin{flushright}
{\sc Pre-Print\ CTP$\sharp$2629}\\
{\sc April 1997}\\
{\tt hep-th/9704206}\\
\end{flushright}
\vspace{.6cm}

\begin{center}
{\Large{\bf Supersymmetric Yang-Mills Theory}}\\[5pt]
{\Large{\bf and}}\\[5pt]
{\Large{\bf Riemannian Geometry
}}\\[40pt]
{\sc Ricardo Schiappa}
\\[5pt]
{\it
Center for Theoretical Physics and Department of Physics\\
Massachusetts Institute of Technology, 77 Massachusetts Avenue\\
Cambridge, MA 02139, U.S.A.}\\[5pt]
{\rm E-mail:} \ {\sc ricardos@mit.edu}\\[40pt]

{\sc Abstract}\\[12pt]

\parbox{13cm}{
We introduce new local gauge invariant variables for $N=1$ supersymmetric 
Yang-Mills theory, explicitly parameterizing the physical Hilbert space of the 
theory. We show that these gauge invariant variables have a geometrical 
interpretation, and can be constructed such that the emergent geometry is 
that of $N=1$ supergravity: a Riemannian geometry with vector-spinor 
generated torsion. Full geometrization of supersymmetric Yang-Mills theory 
is carried out, and geometry independent divergences associated to the 
inversion of a differential operator with zero modes -- that were encountered 
in the non-supersymmetric case -- do not arise in this situation.
}
\end{center}

\vfill

PACS: {\it 02.40.Ky, 11.15.Tk, 11.30.Pb}

Keywords: {\it Supersymmetric Yang-Mills Theory, Gauge Invariance, Geometric 
Variables, Gauss' law}

\end{titlepage}

\sect{Introduction}

For quite sometime now, there exists a nice geometrical setting for 
Yang-Mills theory. That is based on fiber bundle differential geometry, 
where the configuration space is obtained by factoring out time independent 
gauge transformations, and is then seen as the base space of a principal 
fiber bundle, where the structure group is the gauge group \cite{nr}. There 
are many concepts of Riemannian geometry that can then came into the game, 
as there is the possibility of defining a Riemannian metric on the space of 
non-equivalent gauge connections \cite{bv}.

However, this setting must be cast into a more workable form when we want to 
study the strong coupling regime of Yang-Mills theory. In here, gauge 
invariance becomes an important constraint on states of the theory in the 
form of Gauss' law. This constraint amounts to a reduction of the number of 
degrees of freedom present in the gauge connection: if we start with a gauge 
group $G$, in the canonical formalism and in temporal gauge $A^a_0=0$, the 
number of variables is $3\,{\rm dim}\,G$, when in fact we only have $2\,{\rm 
dim}\,G$ physical gauge invariant degrees of freedom. The question of 
whether one can construct local gauge invariant variables is then an 
important one, as it would allow us to easily implement the Gauss' law 
constraint. These variables would then seem the most appropriate ones to 
describe the physical space of the theory. Moreover, observe that in 
temporal gauge the remaining local gauge invariance is now restricted 
to space-dependent transformations at a fixed time. This is the true 
quantum mechanical symmetry of the theory. Working with local gauge 
invariant variables, this symmetry of the Hamiltonian can be maintained 
exactly, even under approximations to the dynamics.

This idea first appeared in \cite{hal} \cite{gj}, and has recently gained 
new momentum with the work in \cite{dzf1} \cite{dzf2} \cite{lun} \cite{kj1} 
\cite{lun2} \cite{ph} \cite{gs} \cite{kj2} \cite{kj3}, and references 
therein. In \cite{kj1}, one constructs a change of variables 
that will allow us to replace the coordinates $A^a_i$ by new coordinates 
$u^a_i$ that have the property of transforming covariantly under the gauge 
group, as opposed to as a gauge connection. Then, in these new coordinates, 
the generator of gauge transformations becomes a (color) rotation generator, 
and by contracting in color we can obtain gauge invariant variables to our 
theory, $g_{ij}\equiv u^a_iu^a_j$. States $\Psi[g_{ij}]$ depending only on 
these gauge invariant variables manifestly satisfy Gauss' law. We must be 
careful, however. Not any choice of gauge covariant variables is adequate: 
an appropriate set of variables should describe the correct number of gauge 
invariant degrees of freedom at each point of space, and should also be 
free of ambiguities such as Wu-Yang ambiguities \cite{wy}. In this case, 
several gauge unrelated vector potentials may lead to the same color 
magnetic field. Variables that are Wu-Yang insensitive are of no use, as in 
the functional integral formulation Wu-Yang related potentials must be 
integrated over -- since they are not gauge related --, while functional 
integration over Wu-Yang insensitive variables always misses these 
configurations. The absence of Wu-Yang ambiguities will be clear if we are 
able to invert the variable transformation, {\it i.e.}, if when 
transforming $A \rightarrow u$ we can also have an explicit expression 
for $A[u]$.

In \cite{kj1}, the set of gauge covariant variables $\{u^a_i\}$ that replace 
the $SU(2)$ gauge connection was defined by the differential equations:
$$
\epsilon^{ijk}D_ju^a_k \equiv \epsilon^{ijk}(\partial_ju^a_k+\epsilon^{abc}
A^b_ju^c_k)=0 \eqno(1.1)
$$
which is equivalent to writing,
$$
\partial_ju^a_k+\epsilon^{abc}A^b_ju^c_k-\Gamma^s_{jk}u^a_s=0 \eqno(1.2)
$$
as the $\{u^a_i\}$ have $\det\,u \not= 0$, and so form a complete basis. 
Observe that we have $\Gamma^s_{jk}=\Gamma^s_{kj}$, and these quantities can 
be written as,
$$
\Gamma^s_{jk}={1 \over 2}g^{sn}(\partial_jg_{nk}+\partial_kg_{jn}-
\partial_ng_{jk}) \eqno(1.3)
$$
where
$$
g_{ij} \equiv u^a_iu^a_j \eqno(1.4)
$$

So, a "metric" tensor was implicitly introduced by the defining equations 
for the new variables, (1.1). Observe that equation (1.2) is simply the 
so-called dreibein postulate, where the $\{u^a_i\}$ plays the role of a 
dreibein, $\omega^{ac}_i \equiv \epsilon^{abc}A^b_i$ is a spin-connection, 
and $\Gamma^i_{jk}$ is the affine metric connection. A torsion-free 
Riemannian geometry in a three manifold was then introduced by the 
definition of the new 
variables. The metric $g_{ij}$ contains in itself the six local gauge 
invariant degrees of freedom of the $SU(2)$ gauge theory. Moreover, any gauge 
invariant wave-functional of $A^a_i$ can be written as a function of 
$g_{ij}$ only, and any wave-functional of $g_{ij}$ is gauge invariant 
\cite{kj1}. This implements gauge invariance exactly. Finally, the dreibein 
postulate can be inverted so that we obtain,
$$
A^a_i=-{1\over2}\epsilon^{abc}u^{bj}\nabla_iu^c_j \eqno(1.5)
$$
where we use the notation $\nabla_ju^a_k \equiv \partial_ju^a_k-
\Gamma^s_{jk}u^a_s$ for the purely geometric covariant derivative 
(as opposed to the gauge covariant derivative). Therefore, the new variables 
avoid Wu-Yang ambiguities.

Full geometrization of Yang-Mills theory in this formulation was then carried 
out in \cite{kj1} \cite{ph}. The electric energy involves the inversion of a 
differential operator that can generically have zero modes. By deforming 
equations (1.1) it was then shown how one could proceed to compute the 
electric tensor \cite{kj2}. Instanton and monopole configurations have been 
identified as the $S^3$ and $S^2 \times {\bf R}$ geometries \cite{kj2}, and, 
more recently, the form the wave-functional for two heavy color sources 
should take has been calculated \cite{kj3}. The computations are carried out 
in the Schroedinger representation of gauge theory, see \cite{j} for a review.

Supersymmetric Yang-Mills theory has also been well established for quite 
sometime now. It allows for many simplifications in quantum computations, 
and with an appropriate choice of matter content and/or number of 
supersymmetry generators, we can obtain finite quantum field theories. 
Textbook references are \cite{bl} \cite{wb} \cite{w}. Moreover, recently 
there has been a lot of progress and activity in the field due to the 
possibility of actually solving for the low-energy effective action of 
certain cases of supersymmetric Yang-Mills theories, starting with the work 
in \cite{sw}. It is then natural to extend the work on gauge invariant 
geometrical variables to the supersymmetric case. That is what we will do here.

We will see that it is possible to define variables that also have a 
geometrical interpretation, namely, as the variables present in supergravity. 
We should point out, however, that no coupling to gravity is ever considered. 
Still, we need a motivation to construct the new variables. As in the pure 
Yang-Mills case the new variables and geometry have an interpretation as 
the variables and geometry of three dimensional gravity, it is natural to 
assume that in the 
supersymmetric case the new variables and geometry could likewise have an 
interpretation as the variables and geometry of three dimensional supergravity. 
This will be 
a guiding principle throughout our work. More geometrical intuition on how 
to construct the new variables will come from an extra symmetry enjoyed by 
both the canonical variables and Gauss' law generator. That is a symmetry 
under $GL(3)$ transformations, a diffeomorphism symmetry. This will allow 
us to naturally assign tensorial properties to diverse local quantities of 
the theory. Obviously the Hamiltonian (or any other global operator) will 
not possess this symmetry. After all, supersymmetric Yang-Mills theory is 
not diffeomorphism invariant.

The plan of this paper is as follows. In section 2 we start by reviewing 
the conventions of $N=1$ supersymmetry, and also outline the geometry of 
supergravity. In section 3 we will then explore the $GL(3)$ symmetry, 
assigning tensorial properties to local (composite) operators. We this in 
hand, we then proceed to define gauge invariant geometrical variables for 
supersymmetric Yang-Mills theory in section 4, carrying out the full 
geometrization of the theory in section 5. Section 6 presents a concluding 
outline.

\sect{Review and Conventions}

The conventions in \cite{bl} \cite{wb} and \cite{w} are basically the same. 
We will follow \cite{bl} with minor changes, as we take $\sigma^{\mu\nu}=
{1\over4}[\gamma^\mu,\gamma^\nu]$. The $N=1$ supersymmetry algebra is 
obtained by introducing one spinor generator, $Q$, which is a Majorana 
spinor, to supplement the usual (bosonic) generators of the Poincar\'e 
group. The $N=1$ supersymmetry algebra is then the Poincar\'e algebra plus:
$$
[P_\mu,Q]=0
$$
$$
[M_{\mu\nu},Q]=-i\sigma_{\mu\nu}Q
$$
$$
\{Q,\bar{Q}\}=2\gamma^\mu P_\mu \eqno(2.1)
$$
where $\bar{Q}\equiv Q\gamma^0$.

Supersymmetric gauge theory, based on gauge group $G$ with gauge algebra 
$\cal G$, has as component fields the gluons, or gauge connection, $A^a_\mu$; 
the gluinos, super-partners of the gauge fields and Majorana spinors, 
$\lambda^a$; and the scalar auxiliary fields $D^a$. All these fields are 
in the adjoint representation of $\cal G$. In Wess-Zumino gauge, the action is,
$$
S=\int d^4x \, \{-{1\over4}F^a_{\mu\nu}F^{a\mu\nu}+{i\over2}\bar{\lambda}^a
\gamma^\mu D_\mu \lambda^a+{1\over2}D^a D^a\} \eqno(2.2)
$$
where we can see that the auxiliary fields have no dynamics. The 
supersymmetry transformation laws of the fields, that leave the action 
invariant are:
$$
\delta A^a_\mu = i \bar{\varepsilon}\gamma_\mu \lambda^a
$$
$$
\delta \lambda^a = (\sigma^{\mu\nu}F^a_{\mu\nu}-i\gamma_5 D^a)\varepsilon
$$
$$
\delta D^a = \bar{\varepsilon}\gamma_5 \gamma^\mu D_\mu \lambda^a \eqno(2.3)
$$
where $\varepsilon$ is a Majorana spinor, which is the parameter of the 
infinitesimal supersymmetry transformation. These transformation laws 
implement a representation of the $N=1$ supersymmetry algebra in the quantum 
gauge field theory. The Noether conserved current of supersymmetry is a 
vector-spinor,
$$
J^\mu = i \gamma^\mu \sigma^{\alpha\beta}F^a_{\alpha\beta} \lambda^a \eqno(2.4)
$$
and so the quantum field theoretic representation of the supersymmetry 
generator is given by the Majorana spinor,
$$
Q=i\int d^3 x \, \gamma^0 \sigma^{\mu\nu}F^a_{\mu\nu} \lambda^a \eqno(2.5)
$$

This outlines our usage of notation for supersymmetric gauge theory. We 
still have to outline notation for the supergravity geometry. In here, we 
will have a graviton, $g_{\mu\nu}$, and a gravitino, which is described by a 
Rarita-Schwinger field, $\psi_\mu$. So, we need to start by reviewing 
notation for inserting spinors in curved manifolds. Having a metric, we can 
define orthonormal frames, and so insert a tetrad base at the tangent space 
to a given point, which will allow us to translate between curved and flat 
indices. In particular, this allows us to introduce gamma matrices in the 
manifold, and so introduce spinors. If we are in a manifold $M$, and pick a 
point $p \in M$, we can introduce a tetrad base $\{u^a_\mu \}$ at $p$ via,
$$
g_{\mu\nu}=u^a_\mu u^b_\nu \eta_{ab} \eqno(2.6)
$$
defining an orthonormal frame at each point on $M$. We can now insert gamma 
matrices as $\gamma^\mu(x)u^a_\mu(x)=\gamma^a$, where the $\gamma^a$ are 
numerical matrices. Local Lorentz transformations in the tangent space 
$T_p M$ are ${\Lambda^a}_b(p)$ and (Dirac) spinors at $p\in M$ rotate as,
$$
\psi_\alpha (p) \rightarrow S_{\alpha\beta}({\Lambda^a}_b(p))\psi_\beta (p) 
\eqno(2.7)
$$

Now, we construct a covariant derivative, ${\bf D}_a\psi_\alpha$, which is a 
local Lorentz vector, and transforms as a spinor,
$$
{\bf D}_a\psi_\alpha \rightarrow S_{\alpha\beta}(\Lambda){\Lambda_a}^b
{\bf D}_b\psi_\beta \eqno(2.8)
$$
That is done via a connection $\Omega_\mu$ such that, ${\bf D}_a\psi=
u_a^\mu(\partial_\mu+\Omega_\mu)\psi$, and,
$$
\Omega_\mu={1\over2}\omega^{ab}_\mu \sigma_{ab}={1\over2}u^a_\nu\nabla_\mu 
u^{b\nu}\sigma_{ab} \eqno(2.9)
$$
where $\omega^{ab}_\mu$ is the spin-connection.

Now that we have spinors defined on curved manifolds, we can proceed with 
supergravity. In here, the Riemannian connection $\Gamma^\rho_{\mu\nu}$ is 
{\it not} torsion-free. It is still metric compatible, so that we can write,
$$
\Gamma^\rho_{\mu\nu}=\hat{\Gamma}^\rho_{\mu\nu}-{K_{\mu\nu}}^\rho \eqno(2.10)
$$
where $\hat{\Gamma}^\rho_{\mu\nu}$ is the affine metric connection, and 
${K_{\mu\nu}}^\rho$ is the contorsion tensor. Hatted symbols will always 
stand for quantities computed via the affine metric connection. The torsion 
tensor is,
$$
{T_{\mu\nu}}^\rho \equiv \Gamma^\rho_{\mu\nu}-\Gamma^\rho_{\nu\mu} \eqno(2.11)
$$
and so,
$$
{K_{\mu\nu}}^\rho = -{1\over2}({T_{\mu\nu}}^\rho - g_{\nu\lambda}
g^{\sigma\rho} {T_{\mu\sigma}}^\lambda - g_{\mu\lambda}g^{\sigma\rho} 
{T_{\nu\sigma}}^\lambda) \eqno(2.12)
$$

In $N=1$ supergravity, the torsion is defined by the Rarita-Schwinger field 
$\psi_\mu$ as,
$$
{T_{\mu\nu}}^a={i\over2}k^2\bar{\psi}_\mu\gamma^a\psi_\nu \eqno(2.13)
$$
where $a$ is a flat index, and $k$ is the gravitational dimensionfull 
constant. The tetrad postulate is,
$$
{\cal D}_\mu u^a_\nu\equiv\partial_\mu u^a_\nu+\omega_\mu^{ab}u_{b\nu}-
\Gamma^\rho_{\mu\nu}u^a_\rho=0 \eqno(2.14)
$$
and the covariant derivative acting on spinor indices is,
$$
({\bf D}_\mu)_{\alpha\beta} \equiv \delta_{\alpha\beta}\partial_\mu 
+{1\over2}\omega_{\mu ab}(\sigma^{ab})_{\alpha\beta} \eqno(2.15)
$$

Finally, the supersymmetry transformations that leave the $N=1$ supergravity 
action (Einstein-Hilbert plus Rarita-Schwinger) invariant are:
$$
\delta u^a_\mu (x) = ik\bar{\xi}(x)\gamma^a\psi_\mu(x)
$$
$$
\delta \psi_\mu(x) = {2\over k}{\bf D}_\mu \xi(x) \eqno(2.16)
$$
where $\xi(x)$ is the infinitesimal parameter of the transformation (now a 
space-time dependent Majorana spinor), and where we have not included the 
auxiliary fields. This ends our review and outline of conventions. We can 
now start analyzing the gauge invariant variables geometrization of 
supersymmetric Yang-Mills theory.

\sect{Canonical Formulation and $GL(3)$ Properties}

In the Lagrangian formulation of the theory, the $N=1$ supersymmetry algebra 
closes only up to the field equations. In order to obtain manifest 
supersymmetry, and off-shell closure of the algebra, we need to introduce 
auxiliary fields. In contrast to this situation, it is known that in the 
canonical formalism the $N=1$ super Lie algebra closes without the 
introduction of auxiliary fields (in terms of Dirac brackets the algebra 
closes strongly; otherwise it closes weakly, {\it i.e.}, up to the 
first-class constraints) \cite{ds1} \cite{ds2}. So, we drop the auxiliary 
fields.

The Hamiltonian for supersymmetric gauge theory is therefore,
$$
H=\int d^3x \{{1\over2}e^2(E^{ai})^2+{1\over{2e^2}}(B^{ai}[A^b_j])^2-
{i\over2}\bar{\lambda}^a\gamma^iD_i\lambda^a\} \eqno(3.1)
$$
where $e$ is the coupling constant. The gauge covariant derivative is,
$$
D_i\lambda^a=\partial_i\lambda^a+f^{abc}A^b_i\lambda^c \eqno(3.2)
$$
and the magnetic field potential energy,
$$
B^{ai}[A^b_n]\equiv{1\over2}\epsilon^{ijk}F^a_{jk}[A^b_n]=
\epsilon^{ijk}(\partial_jA^a_k+{1\over2}f^{abc}A^b_jA^c_k) \eqno(3.3)
$$

We still have to impose the Gauss' law constraint on the physical states of 
the theory,
$$
{\cal G}^a(x) \equiv D_iE^{ai}(x)-{i\over2}f^{abc}\lambda^b(x)
\lambda^c(x) \ \ \ , \ \ \ {\cal G}^a(x)\Psi[A^b_i,\lambda^c]=0 \eqno(3.4)
$$
This local composite operator is the generator of local gauge transformations.

There is one more element in the $N=1$ supersymmetric Yang-Mills theory, and 
that is the Majorana spinor $Q$, the generator of supersymmetry. Using the 
definition,
$$
Q\equiv\int d^3x \, {\cal Q}(x) \eqno(3.5)
$$
we can then write,
$$
{\cal Q}(x)=i(-e\gamma_iE^{ai}(x)+{1\over e}\epsilon_{ijk}\gamma^0
\sigma^{ij}B^{ak}[A^b_n(x)])\lambda^a(x) \eqno(3.6)
$$
or, using the explicit Weyl representation of the gamma matrices, we can 
equivalently write this local composite operator in a more compact form,
$$
{\cal Q}(x)=i\pmatrix{
0 & (eE^{ai}(x)+{i\over e}B^{ai}[A^b_n(x)])\sigma_i \cr
(-eE^{ai}(x)+{i\over e}B^{ai}[A^b_n(x)])\sigma_i & 0 \cr
}\lambda^a(x) \eqno(3.7)
$$

Now, in the bosonic half of the theory, the canonical variables are 
$A^a_i(x)$ and $E^{ai}(x)$. Canonical quantization is carried out by the 
commutator,
$$
[A^a_i(x),E^{bj}(y)]=i\delta^{ab}\delta^i_j\delta(x-y) \eqno(3.8)
$$
The momentum $E^{ai}(x)$ will be implemented as a functional derivative 
acting on wave-functionals as,
$$
E^{ai}(x)\Psi[A^b_n,\lambda^c] \rightarrow -i{\delta \over 
{\delta A^a_i(x)}}\Psi[A^b_n,\lambda^c] \eqno(3.9)
$$
In the fermionic half of the theory, we have a Majorana spinor 
$\lambda^a(x)$. Canonical quantization is carried out by establishing the 
anti-commutation relations for the spinorial field,
$$
\{\lambda^a_\alpha(x),\lambda^b_\beta(y)\}=\delta^{ab}\delta_{\alpha\beta}
\delta(x-y) \eqno(3.10)
$$
Both the commutator and the anti-commutator are to be evaluated at equal 
times. We can now compute the commutators and anti-commutators of this 
theory, which involve the composite operators $H$, ${\cal G}^a(x)$ and 
${\cal Q}(x)$. Clearly, these (anti)-commutators are related to the 
symmetry transformations generated by these operators.

The commutators involving the generator of local gauge transformations of 
the canonical variables can be computed to be,
$$
[A^a_i(x),{\cal G}^b(y)]=i(\delta^{ab}\partial_i-f^{acb}A^c_i(x))\delta(x-y) 
\eqno(3.11)
$$
$$
[E^{ai}(x),{\cal G}^b(y)]=if^{abc}E^{ci}(x)\delta(x-y) \eqno(3.12)
$$
$$
[\lambda^a(x),{\cal G}^b(y)]=if^{abc}\lambda^c(x)\delta(x-y) \eqno(3.13)
$$
and the (anti)-commutators involving the local composite operator associated 
to the supersymmetry generator can similarly be found to be,
$$
[A^a_i(x),{\cal Q}(y)]=-\gamma_i\lambda^a(x)\delta(x-y) \eqno(3.14)
$$
$$
[E^{ai}(x),{\cal Q}(y)]=\epsilon^{ijk}(\epsilon_{knm}\gamma^0\sigma^{nm})
D_j\lambda^a\delta(x-y) \eqno(3.15)
$$
$$
\{\lambda^a(x),{\cal Q}(y)\}=i(-e\gamma_iE^{ai}(x)+{1\over e}\epsilon_{ijk}
\gamma^0\sigma^{ij}B^{ak}[A^b_n(x)])\delta(x-y) \eqno(3.16)
$$

Moreover, we will also have that the Hamiltonian and the supersymmetry 
generator are both gauge invariant composite operators, as,
$$
[H,{\cal G}^a(x)]=0 \eqno(3.17)
$$
$$
[{\cal Q}(x),{\cal G}^a(y)]=0 \eqno(3.18)
$$

As expected, the generators ${\cal G}^a(x)$ define the local gauge algebra,
$$
[{\cal G}^a(x),{\cal G}^b(y)]=if^{abc}{\cal G}^c(x)\delta(x-y) \eqno(3.19)
$$
The supersymmetry generator $Q$ defines, along with the generators of the 
Poincar\'e algebra, the $N=1$ supersymmetry algebra \cite{ds1}. However, the 
defined local composite operator ${\cal Q}(x)$ does not define a local 
algebra. That is to be expected as we do not have local supersymmetry in 
our theory. This local operator was only introduced in order to facilitate 
the following tensorial analysis based on diffeomorphism transformations of 
the presented (anti)-commutators.

So, we now want to check that there is a $GL(3)$ symmetry at work for the 
formulae (3.8), (3.10) and (3.11-13), (3.19). The bosonic part tensorial 
assignments will be just like in the pure Yang-Mills case \cite{kj1}, as is 
to be expected. So, the mentioned canonical relations are covariant under 
diffeomorphisms $x^i \rightarrow y^n(x^i)$ on the domain ${\bf R}^3$, 
provided $A^a_i(x)$ is a one-form in ${\bf R}^3$, transforming as
$$
{A'}^a_n(y^m)={{\partial x^i}\over{\partial y^n}}A^a_i(x^j) \eqno(3.20)
$$
where $[\partial x^i/\partial y^n]$ is a $GL(3)$ matrix. That 
${\bf A}(x)\equiv A^a_\mu(x)T^adx^\mu$ is a Lie algebra valued one-form is a 
well known fact from the fiber bundle geometry of gauge theory; so 
consistency holds. Also, provided $E^{ai}(x)$ is a vector density (weight 
$-1$) in ${\bf R}^3$, transforming as
$$
{E'}^{an}(y^m)=\det [{{\partial x^i}\over{\partial y^n}}]{{\partial y^n}\over
{\partial x^i}}E^{ai}(x^j) \eqno(3.21)
$$
The same property holds for $B^{ak}(x)$. This is consistent with the 
implementation of $E^{ai}(x)$ as a functional derivative (3.9), and with the 
definition of the magnetic field (3.3). Commutator (3.8) is then clearly 
diffeomorphic invariant, without the intervention of a space metric. However, 
to introduce spinors, we do need a metric (more precisely, a dreibein base). 
We will assume we do have a metric, $g_{ij}$, and later we will construct it 
using the bosonic dynamical variables of the theory.

When restricted to three dimensional Euclidean space, Lorentz 
transformations become rotations in ${\bf R}^3$. The spinor representation 
of a rotation is then, at a point $p \in M$, given by the orthogonal matrix 
acting on spinor indices,
$$
S_{\alpha\beta}(\Lambda(p))=\exp({1\over2}\omega_{ab}(\Lambda(p))
\sigma^{ab})_{\alpha\beta} \eqno(3.22)
$$
where $\omega$ is the rotation parameter. We can now define the $GL(3)$ 
properties of $\lambda^a(x)$, in order to maintain the anti-commutator 
(3.10) diffeomorphism invariant. That relation is invariant under 
diffeomorphisms, provided $\lambda^a(x)$ is a spinorial density (weight 
$-{1\over2}$) in ${\bf R}^3$, transforming as
$$
{\lambda'}^a_\alpha(y^m)=\det[{{\partial x^i}\over{\partial y^n}}]^{1\over2}
S_{\alpha\beta}(\Lambda(p))\lambda^a_\beta(x^j) \eqno(3.23)
$$

Let us now see what are the consequences of these $GL(3)$ properties on the 
composite local operators ${\cal G}^a(x)$ and ${\cal Q}(x)$. Starting with 
the generator of local gauge transformations, we can observe that the 
tensorial properties of the canonical variables imply that under 
diffeomorphisms we will have that ${\cal G}^a(x)$ is a scalar density 
(weight $-1$) in ${\bf R}^3$, transforming as
$$
{{\cal G}'}^a(y^m)=\det[{{\partial x^i}\over{\partial y^n}}]{\cal G}^a(x^j) 
\eqno(3.24)
$$
This automatically verifies that the canonical commutators (3.11-13) and 
(3.19) are invariant under local diffeomorphisms on the domain of the local 
canonical variables.

Now, look at the other local composite operator, ${\cal Q}(x)$, (3.6) or 
(3.7). First observe that in (3.7) the Pauli matrices $\vec\sigma$ are 
numerical matrices, and not dynamical ones (in which case we would write 
$\sigma^i(x)u^a_i(x)=\sigma^a$, $\{u^a_i\}$ being a dreibein base). We 
write (3.7) as:
$$
{\cal Q}(x) \equiv \sigma_i \otimes {\cal q}^i(x) \eqno(3.25)
$$
where,
$$
{\cal q}^i(x) \equiv i \pmatrix{
0 & eE^{ai}(x)+{i \over e}B^{ai}[A^b_n(x)] \cr
-eE^{ai}(x)+{i \over e}B^{ai}[A^b_n(x)] & 0 \cr
}\lambda^a(x) \eqno(3.26)
$$

Then, the tensorial properties under $GL(3)$ of the canonical variables 
imply that, under diffeomorphisms, we will have that ${\cal q}^i(x)$ is a 
vector-spinor density (weight $-{3\over2}$) in ${\bf R}^3$, transforming as
$$
{{\cal q}'}^n_\alpha(y^m)=\det[{{\partial x^i}\over{\partial y^n}}]^{3\over2}
{{\partial y^n}\over{\partial x^i}}S_{\alpha\beta}(\Lambda(p))
{\cal q}^i_\beta(x^j) \eqno(3.27)
$$
However, as the $\sigma_i$'s in (3.25) are numerical, they do not transform 
under the diffeomorphism, and so ${\cal Q}(x)$ fails to be covariant. This 
is to be expected, as we will see below.

So, the $GL(3)$ symmetry of the (anti)-commutation relations involving 
local (composite) operators and local variables has been established, given 
the tensorial properties assigned to the canonical variables. Clearly, the 
theory itself fails to be $GL(3)$ invariant, and that is to be expected: the 
Hamiltonian is not covariant under diffeomorphisms (the metric $\delta_{ij}$ 
appears instead of $g_{ij}$, the measure $d^3x$ appears instead of 
$\sqrt{g}d^3x$, {\it etc.}). This can be related to the lack of covariance of 
the supersymmetry generator (3.25), (3.27). Indeed, we can see ${\cal Q}(x)$ 
as the square root of the Hamiltonian; so if the Hamiltonian fails to be 
covariant, so should the supersymmetry generator. Moreover, observe that 
when we square (3.25) we will obtain a term like $\sigma_i\sigma_j=
\delta_{ij}+i\epsilon_{ijk}\sigma_k$, and this can be seen as the origin of 
the "wrong" metric $\delta_{ij}$ in the Hamiltonian, which will destroy the 
possibility of local covariance. Also, no global (composite) operator can 
have this $GL(3)$ symmetry, due to the "wrong" choice of integration 
measure. Now that we have assigned tensorial properties to local quantities 
in our supersymmetric theory, we are ready to proceed in looking for 
further geometrization in this canonical framework.

\sect{Geometric Variables}

We will now limit ourselves to the simplest case of non-abelian gauge 
group, namely, $G=SU(2)$. Then, the structure constants are simply 
$f^{abc}=\epsilon^{abc}$. We will assume knowledge of the previous work 
done for pure Yang-Mills theory \cite{kj1} \cite{ph} \cite{kj2}.

We will want to have a representation of supersymmetry once we are in the 
new variables. As we know from the bosonic case \cite{kj1}, the gluon field 
is transformed into a "metric" field. The supersymmetry representation that 
includes a metric field is that of supergravity, and it also includes a 
vector-spinor field. So, we will expect that the gluino field will be 
transformed into a "gravitino" field. We will therefore want to transform 
the supersymmetric Yang-Mills variables, $\{A^a_i(x),\lambda^a(x)\}$, into 
the variables of three dimensional supergravity, $\{g_{ij}(x),\psi_k(x)\}$. 
We will also expect to obtain a geometry similar to the one of supergravity. 
After all, the defining equation for the $\{u^a_i(x)\}$ variables 
identifies them with a dreibein base in a three dimensional manifold.

Recall form section 2 what we are to expect. The geometry will have 
torsion, defined as,
$$
{T_{ij}}^a={i\over2}\bar{\psi}_i\gamma^a\psi_j \eqno(4.1)
$$
We can insert a dreibein base through,
$$
g_{ij}=u^a_iu^b_j\delta^{ab} \eqno(4.2)
$$
and we also expect that there could be some local supersymmetry 
transformation in these new variables, which we will call the "geometrical 
supersymmetry variation", and which would look like the supersymmetric 
transformation laws of supergravity,
$$
\delta u^a_i(x)=i\bar{\xi}(x)\gamma^a\psi_i(x)
$$
$$
\delta\psi_k(x)=2{\bf D}_k\xi(x) \eqno(4.3)
$$
where the covariant derivative acting on spinor indices was defined in 
(2.15). With this at hand, the dreibein postulate is now written as,
$$
{\cal D}_ju^a_k \equiv \partial_ju^a_k+\omega^{ab}_ju^b_k-\Gamma^n_{jk}u^a_n
=0 \eqno(4.4)
$$
also defining the symbol ${\cal D}_j$. Multiplying this equation by 
$\epsilon^{ijk}$, and defining the spin-connection via the gauge 
connection, as in,
$$
\omega^{ab}_j(x) \equiv \epsilon^{acb}A^c_j(x) \eqno(4.5)
$$
the dreibein postulate becomes,
$$
\epsilon^{ijk}D_ju^a_k=\epsilon^{ijk}(\partial_ju^a_k+\epsilon^{abc}
A^b_ju^c_k)={1\over2}\epsilon^{ijk}{T_{jk}}^a={i\over4}\epsilon^{ijk}
\bar{\psi}_j\gamma^a\psi_k \eqno(4.6)
$$

We will take these differential equations to define the change of 
variables $A^a_i(x) \rightarrow u^a_i(x)$. Then, the reverse line of 
argument holds: the new variables $\{u^a_i(x)\}$ play the role of a 
dreibein, and from them we can construct a metric $g_{ij}=u^a_iu^a_j$ 
which is a local gauge invariant variable. The geometry defined by this 
new variable has torsion, given by (4.1). Clearly, for the change of 
variables to be well defined, we still need to specify what $\psi_j(x)$ is. 
That is the problem we will now address.

We begin with some dimensional analysis. We know that the gauge field 
$A^a_i(x)$ has mass dimension one, and the gaugino field $\lambda^a(x)$ has 
mass dimension three halfs. We also know that the mass dimension of the 
fermionic generator of the supersymmetry algebra is one half. Through 
definition (3.5) and expression (3.25) we observe that if we ever were to 
modify the gauge theory in order to covariantize it (inserting 
$\sigma^i(x)u^a_i(x)=\sigma^a$ in (3.25) and from then on), we would need 
the dreibein field to have zero mass dimension, as well as the metric. 
Though we are not going to modify the gauge theory in this work, we may as 
well stick to this broader perspective. Then, through the dreibein defining 
equation (4.6), we conclude that the "gravitino" field has mass dimension 
one half. These dimensional assignments are just like what happens in 
supergravity.

We can now see that this will have some influence on the construction of 
the "gravitino" defining equation. In fact, there are some {\it a priori} 
requirements for such an equation. It must be geometrical, either in a 
differential or algebraic way; we need 12 equations, to change the 12 
variables $\lambda^a_\alpha$ to the 12 variables $\psi_{k\alpha}$; and the 
gluino field must be present in such an equation. If we moreover require 
linearity on fermionic variables (like we had linearity on the bosonic 
variables in (4.6)), we see that, by simple dimensional analysis, we can not 
write such an equation algebraically, but only differentially. Moreover, 
the equation is constrained to be of the form,
$$
\epsilon^{ijk}{\bf D}_j\psi_{k\alpha}={\cal M}^{ia}_{\alpha\beta}\lambda^a_
\beta \eqno(4.7)
$$
where the matrix ${\cal M}^{ia}_{\alpha\beta}$ must have zero mass 
dimension, being so far otherwise arbitrary. However, we must be careful. 
Not only do we want to have a geometrical way in which to define the 
vector-spinor field, but we also want to be compatible with the fact that 
we are studying a supersymmetric theory. In particular, we would like the 
geometrical supersymmetry variation (4.3) to generate the gauge 
supersymmetry variation (2.3). So we will ask for the geometrical 
variation (4.3) to generate the gauge supersymmetry variation on the 
bosonic variables $A^a_i(x)$, and in the simplest case where 
$\xi(x)\equiv\varepsilon$.

Under a generic variation of the fields, we obtain for (4.6),
$$
\epsilon^{ijk}D_j\delta u^a_k=-\epsilon^{ijk}\epsilon^{abc}u^c_k
\delta A^b_j+{1\over2}\epsilon^{ijk}\delta {T_{jk}}^a \eqno(4.8)
$$
where,
$$
{1\over2}\epsilon^{ijk}\delta{T_{jk}}^a={i\over2}\epsilon^{ijk}
\bar{\psi}_j\gamma^a\delta\psi_k \eqno(4.9)
$$
The supersymmetry transformation laws we will need are (2.3) and (4.3). 
So, a supersymmetry transformation of the dreibein defining equation yields 
the "gravitino" defining equation. Performing the computations, based on the 
previous formulae, we are led to,
$$
\epsilon^{ijk}{\bf D}_j\psi_k=\epsilon^{ijk}(\partial_j\psi_k+{1\over2}
\omega_j^{ab}\sigma^{ab}\psi_k)={1\over3}\epsilon^{ijk}\epsilon^{abc}
\gamma^a\gamma_j\lambda^bu^c_k \eqno(4.10)
$$
where $\gamma_i(x)=u^a_i(x)\gamma^a$.

We will take these differential equations to define the change of 
variables $\lambda^a(x) \rightarrow \psi_k(x)$. Observe that this equation 
is precisely of the required form (4.7), and the matrix 
${\cal M}^{ia}_{\alpha\beta}$ has been uniquely defined. Also, this alone 
guarantees that the geometric variation (4.3) will generate the bosonic 
gauge supersymmetry variation (2.3), when $\xi\equiv\varepsilon$. This 
does not guarantee however that the geometric variation will generate the 
fermionic gauge supersymmetry variation under the same circumstances. 
In fact, we can choose $\xi\equiv\xi[\varepsilon]$ through a differential 
equation (4.20) for $\xi$, such that the geometric variation generates the 
gauge supersymmetry variation on $\lambda^a(x)$, but we will not have 
$\xi\equiv\varepsilon$ in this case. This shows that even though we can 
generate the gauge supersymmetry variation via the geometrical 
supersymmetry variation under special circumstances, the geometric 
variation is not the original supersymmetry of Yang-Mills theory. The 
actual expressions for the supersymmetry variations on the new geometrical 
variables can nevertheless be computed using the usual expression,
$$
\delta \Phi = i[Q,\Phi] \eqno(4.11)
$$
where $\Phi$ is any of the geometrical variables, and where we should 
express the supersymmetry generator in this geometric framework (see 
section 5). The resulting expressions would not be as simple as (2.3) or (4.3).

All together, we see that we can now define local gauge invariant geometric 
variables for supersymmetric Yang-Mills theory via the system of coupled 
non-linear partial differential equations, (4.6) and (4.10). These equations 
define a variable change $\{A^a_i,\lambda^b\}\rightarrow\{u^a_i,\psi_k\}$. 
They also introduce a three dimensional Riemannian geometry with torsion 
as given by (4.1-2) and (4.4).

Now that the definition of the new geometrical gauge invariant variables is 
concluded, we would like to invert the defining equations, in order to 
express $A^a_i(x)$ and $\lambda^a(x)$ in terms of the geometric variables. 
This inversion will make clear that there are no Wu-Yang ambiguities related 
to these new variables. The defining equation for the dreibein (4.6) is 
equivalent to the dreibein postulate (4.4), where the connection is with 
torsion,
$$
\Gamma^n_{jk}=\hat{\Gamma}^n_{jk}-{K_{jk}}^n \eqno(4.12)
$$
hatted symbols always denoting affine metric connection quantities. The 
contorsion tensor is computed from the torsion tensor, through (2.12), and 
we obtain,
$$
K_{ijn}={i\over4}(\bar{\psi}_i\gamma_j\psi_n+\bar{\psi}_j\gamma_i\psi_n-
\bar{\psi}_i\gamma_n\psi_j) \eqno(4.13)
$$
Define a purely geometric derivative through,
$$
\nabla_ju^a_k\equiv\partial_ju^a_k-\Gamma^n_{jk}u^a_n \eqno(4.14)
$$
and we can find the expression for the inversion,
$$
A^a_i(x)=-{1\over2}\epsilon^{abc}u^{bk}(x)\nabla_iu^c_k(x) \eqno(4.15)
$$

We will now compute a generic variation of this equation, so that we can 
later use it to compute the inversion for the gluino field. In order to 
carry out the calculation, we will need to know what is the generic 
variation of the connection (4.12). Using the fact that it is metric 
compatible, this can be computed to be,
$$
\delta\Gamma^n_{jk}={1\over2}g^{nm}(\nabla_j\delta g_{mk}+\nabla_k\delta 
g_{mj}-\nabla_m\delta g_{jk})-\delta{K_{jk}}^n \eqno(4.16)
$$
We can now carry out the variation of the dreibein postulate, and from there 
obtain the variation of equation (4.15). The result will be,
$$
\delta A^a_i={\epsilon^{nml}\over{2\sqrt{g}}}u^a_m(\nabla_i(u^b_l\delta 
u^b_n)+\nabla_l(\delta g_{ni})+{i\over2}((\bar{\psi}_i\gamma^b\psi_l)\delta 
u^b_n+{1\over2}(\bar{\psi}_n\gamma^b\psi_l)\delta u^b_i)+
$$
$$
+{i\over2}(\bar{\psi}_n\gamma_l\delta\psi_i+\bar{\psi}_i\gamma_n\delta
\psi_l+\bar{\psi}_n\gamma_i\delta\psi_l)) \eqno(4.17)
$$
where $\sqrt{g}=\det u$.

We will now use this equation to invert for the gluino field. As we know 
the geometrical variation (4.3) with $\xi\equiv\varepsilon$ generates the 
bosonic supersymmetry variation. So, we only need to use (2.3) and (4.3) in 
(4.17), and rearrange, so that we find the expression for the inversion,
$$
\lambda^a(x)=-{\epsilon^{nml}\over{6\sqrt{g(x)}}}u^a_m(x)\gamma^i(x)(
\gamma_l(x){\cal D}_i\psi_n(x)+\gamma_n(x){\cal D}_l\psi_i(x)+\gamma_i(x)
{\cal D}_l\psi_n(x)) \eqno(4.18)
$$
where we use the vector-spinor full covariant derivative, defined as,
$$
{\cal D}_i\psi_{k\alpha}\equiv\partial_i\psi_{k\alpha}+{1\over2}
\omega^{ab}_i(\sigma^{ab})_{\alpha\beta}\psi_{k\beta}-\Gamma^s_{ik}
\psi_{s\alpha} \eqno(4.19)
$$
Observe that even though the spin-connection is defined via the gauge 
connection, it is a fully geometric quantity through the dreibein postulate. 
Later on we will also require an expression for the generic variation of 
this equation, so we will address such a problem now. The computation is 
rather long, and so is the result. We will obtain,
$$
\delta\lambda^a=-{\epsilon^{nml}\over{6\sqrt{g}}}u^a_m(u^{bi}\delta u^c_l+
u^c_l\delta u^{bi})\gamma^b\gamma^c({\cal D}_i\psi_n-{\cal D}_n\psi_i)-
$$
$$
-{\epsilon^{nml}\over{6\sqrt{g}}}u^a_m\gamma^i(\gamma_l{\cal D}_i(
\delta\psi_n)+\gamma_n{\cal D}_l(\delta \psi_i)+\gamma_i{\cal D}_l(
\delta\psi_n)-
$$
$$
-{1\over2}(\gamma_l\sigma^{jk}\psi_n\nabla_i+\gamma_n\sigma^{jk}\psi_i
\nabla_l+\gamma_i\sigma^{jk}\psi_n\nabla_l)(u^b_j\delta u^b_k)-
$$
$$
-{1\over2}(\gamma_l\sigma^{jk}\psi_n\nabla_j(\delta g_{ki})+\gamma_n
\sigma^{jk}\psi_i\nabla_j(\delta g_{kl})+\gamma_i\sigma^{jk}\psi_n
\nabla_j(\delta g_{kl}))+
$$
$$
+{1\over2}(\gamma_l\sigma^{jk}\psi_n\delta K_{ijk}+\gamma_n\sigma^{jk}
\psi_i\delta K_{ljk}+\gamma_i\sigma^{jk}\psi_n\delta K_{ljk})+(\gamma_l
\delta{K_{in}}^s+\gamma_n\delta{K_{li}}^s+\gamma_i\delta{K_{ln}}^s)\psi_s)-
$$
$$
-{1\over{6\sqrt{g}}}u^a_ju^c_k(\epsilon^{njk}\delta u^{cl}-\epsilon^{ljk}
\delta u^{cn})\gamma^i(\gamma_l{\cal D}_i\psi_n+\gamma_n{\cal D}_l\psi_i+
\gamma_i{\cal D}_l\psi_n) \eqno(4.20)
$$
where the generic variation of the contorsion tensor can be written as,
$$
\delta K_{inl}={i\over4}((\bar{\psi}_i\gamma^a\psi_l)\delta u^a_n+
(\bar{\psi}_n\gamma^a\psi_l)\delta u^a_i-(\bar{\psi}_i\gamma^a\psi_n)
\delta u^a_l)+
$$
$$
+{i\over4}((\bar{\psi}_n\gamma_l-\bar{\psi}_l\gamma_n)\delta\psi_i+
(\bar{\psi}_i\gamma_n+\bar{\psi}_n\gamma_i)\delta\psi_l-(\bar{\psi}_l
\gamma_i+\bar{\psi}_i\gamma_l)\delta\psi_n) \eqno(4.21)
$$
The variations (4.17) and (4.20) allow us now to express a variation of the 
wave-functional in terms of the variations of the geometric variables. This 
will be helpful in section 5.

The inversion completed proves the non existence of Wu-Yang ambiguities in 
the new geometrical variables. Therefore, we have managed to define new 
gauge invariant variables for supersymmetric Yang-Mills. Moreover, it can 
be shown that gauge invariant physical wave-functionals of the theory 
depend only on these geometric variables (see section 5), 
$\Psi\equiv\Psi[g_{ij},\psi_k]$, so that we have in these variables an 
explicit parameterization of the physical Hilbert space (moduli space) of 
the gauge theory. A final remark on diffeomorphisms is now in order. As 
said before, only the variables of the theory are diffeomorphism covariant. 
The Hamiltonian fails to be diffeomorphism covariant. Given that the 
variables of the theory are now $\{g_{ij},\psi_k\}$, this has an 
interesting consequence: a configuration diffeomorphic to the previous 
one yields a different configuration to the gauge theory. Therefore, we 
can extend solutions to the gauge theory by action of the group of 
diffeomorphisms, by simply moving along the orbit of the geometrical 
configuration.

\sect{Gauge Tensors as Geometric Tensors}

We now wish to write the tensors and composite operators of our theory in 
terms of the new geometric variables, {\it i.e.}, as geometric tensors and 
geometric composite operators. We will first address the electric and 
magnetic tensors. The Hamiltonian, Gauss' law generator, and the 
supersymmetry generator composite operators then easily follow from these 
two tensors and the previous equations for the inversions of the gluon and 
gluino fields.

Let us start with the gauge Ricci identity,
$$
\epsilon^{abc}F^b_{ij}=[D_i,D_j]^{ac} \eqno(5.1)
$$
and apply it to the dreibein field. We will obtain,
$$
\epsilon^{abc}F^b_{ij}u^c_k={R^l}_{kij}u^a_l \eqno(5.2)
$$
where ${R^l}_{kij}$ is the Riemann tensor of the connection $\Gamma$,
$$
{R^l}_{kij}=\partial_i\Gamma^l_{jk}-\partial_j\Gamma^l_{ik}+\Gamma^m_{jk}
\Gamma^l_{im}-\Gamma^m_{ik}\Gamma^l_{jm} \eqno(5.3)
$$
Now from (5.2) we can express the field strength in terms of the Riemann 
curvature, and so from (3.3) we can express the magnetic field vector 
geometrically, as,
$$
B^{am}=-{1\over{4\sqrt{g}}}\epsilon^{mij}\epsilon^{nlk}u^a_nR_{lkij} \eqno(5.4)
$$
So, the gauge invariant tensor which gives the Yang-Mills magnetic energy 
density is,
$$
B^{ai}B^{aj}={1\over16}\epsilon^{imn}\epsilon^{jkl}{R^{uv}}_{mn}(R_{uvkl}-
R_{vukl}) \eqno(5.5)
$$
As we can see, this expression gives the gauge invariant tensor in a manifestly 
gauge invariant form, in terms of the "metric" $g_{ij}$, and the "gravitino" 
$\psi_k$ (which is present via the torsion contribution to the Riemann tensor).

The electric field vector is the momentum canonically conjugated to the 
canonical variable, the gauge connection. In canonical quantization it is 
represented by a functional derivative (3.9). We will define a gauge 
invariant tensor operator $e^{ij}$ by,
$$
{\delta\over{\delta A^a_i(x)}}=iE^{ai}(x) \equiv \sqrt{g(x)}u^a_j(x)
e^{ij}(x) \eqno(5.6)
$$
Clearly, $e^{ij}(x)$ is an ordinary $(^2_0)$ tensor under $GL(3)$. From this 
expression, the electric gauge invariant Yang-Mills tensor, {\it i.e.}, the 
manifestly gauge invariant tensor which gives the Yang-Mills electric energy 
density, now follows as,
$$
E^{ai}E^{aj}=-g{e^i}_ke^{jk} \eqno(5.7)
$$

In order to finally obtain the Hamiltonian in a manifestly gauge invariant 
form in terms of the geometrical variables, we still need the expression for 
the fermionic energy density, as is clear from (3.1). The expression for this 
gauge invariant tensor can be obtained by simply inserting (4.18-19) in the 
required expression. The result we will obtain is,
$$
\bar{\lambda}^a\gamma^iD_i\lambda^a={{\epsilon^{njk}}\over{36\sqrt{g}}}g_{jm}
((\bar{{\cal D}}_k\bar{\psi}_n)\gamma_l+(\bar{{\cal D}}_l\bar{\psi}_n)\gamma_k+
$$
$$
+(\bar{{\cal D}}_k\bar{\psi}_l)\gamma_n)\gamma^l\gamma^i\nabla_i(
{\epsilon^{umv}\over\sqrt{g}}\gamma^s(\gamma_v({\cal D}_s\psi_u)+\gamma_u
({\cal D}_v\psi_s)+\gamma_s({\cal D}_v\psi_u))) \eqno(5.8)
$$
where we have defined $\bar{{\cal D}}_i\bar{\psi_j} \equiv ({\cal D}_i\psi_j)
^{\dagger}\gamma^0$; and where in the contraction $\gamma^i\nabla_i$ the 
gamma matrices are to be considered as numerical, not as space dependent. The 
sum of (5.5), (5.7) and (5.8) according to (3.1) finally yields the 
manifestly gauge invariant Hamiltonian.

As was done for the gluon functional derivative, we will now similarly 
define a gauge invariant vector-spinor density operator 
$\chi_i$ to deal with the gluino functional derivative,
$$
{\delta\over{\delta\lambda^a(x)}}\equiv\sqrt{g(x)}u^{ai}(x)\chi_i(x) \eqno(5.9)
$$
$\chi_i(x)$ is a $(^0_1)$ vector-spinor density (weight ${1\over2}$) 
under $GL(3)$. With these definitions at hand, we can now express the 
functional dependence of the wave-functional $\Psi[A^a_i,\lambda^b]$ in 
terms of the new variables. Under a variation, we will have,
$$
\delta\Psi=\int d^3x \, \{ {{\delta\Psi}\over{\delta A^a_i(x)}}\delta 
A^a_i(x)+{{\delta\Psi}\over{\delta\lambda^a(x)}}\delta\lambda^a(x) \} =
$$
$$
=\int d^3x \, \{ \sqrt{g(x)}u^a_j(x)\delta A^a_i(x)[e^{ij}(x)\Psi]+
\sqrt{g(x)}u^{ai}(x)[\chi_i(x)\Psi]\delta\lambda^a(x) \} \eqno(5.10)
$$
where we should use the expression for the variations of the gauge fields 
in (4.17) and (4.20-21). Expanding this expression through rather lengthy 
calculations, it can then be seen that the term in $\delta u^a_i$ is 
proportional to the Gauss' law operator (3.4), when expressed in 
geometrical terms, and acting on the wave-functional. Observe that this 
is ${\cal G}^a\Psi[g,\psi]=0$, which in the new variables can be 
written as,
$$
(\nabla_ie^{ij}+{i\over2}\bar{\psi}_i\gamma^k\psi_ke^{ij}+{1\over{72g^2}}(
g^{jn}\epsilon^{ilk}-g^{jl}\epsilon^{ink})((\bar{{\cal D}}_s\bar{\psi}_n)
\gamma_l+(\bar{{\cal D}}_l\bar{\psi}_s)\gamma_n+
$$
$$
+(\bar{{\cal D}}_l\bar{\psi}_n)
\gamma_s)\gamma^s\gamma^0\gamma^r(\gamma_k({\cal D}_r\psi_i)+\gamma_i({\cal D}
_k\psi_r)+\gamma_r({\cal D}_k\psi_i)))\Psi[g,\psi]=0 \eqno(5.11)
$$
So, wave-functionals 
whose dependence is solely on the new gauge invariant variables are gauge 
invariant, and gauge invariant wave-functionals depend solely on the new 
gauge invariant variables. It is in these physical gauge invariant 
wave-functionals that we are mainly interested, and for these the 
previous expression for $\delta\Psi$ reduces to,
$$
\delta\Psi[g_{ij},\psi_k] = \int d^3x \, \{ {1\over2}\epsilon^{nml}
\nabla_l(\delta g_{ni})[e^i_m\Psi]+
$$
$$
+{1\over{12}}\epsilon^{nml}[\chi_m\Psi]\gamma^i(\gamma_l\sigma^{jk}\psi_n
\nabla_j(\delta g_{ki})+\gamma_n\sigma^{jk}\psi_i\nabla_j(\delta g_{kl})+
\gamma_i\sigma^{jk}\psi_n\nabla_j(\delta g_{kl})) +
$$
$$
+{i\over4}\epsilon^{nml}(\bar{\psi}_n\gamma_l\delta\psi_i+\bar{\psi}_i
\gamma_n\delta\psi_l+\bar{\psi}_n\gamma_i\delta\psi_l)[e^i_m\Psi]-
$$
$$
-{1\over6}\epsilon^{nml}[\chi_m\Psi]\gamma^i(\gamma_l{\cal D}_i(
\delta\psi_n)+\gamma_n{\cal D}_l(\delta\psi_i)+\gamma_i{\cal D}_l(
\delta\psi_n))-
$$
$$
-{i\over{24}}\epsilon^{nml}[\chi_m\Psi]\gamma^i(\gamma_l\sigma^{jk}
\psi_n(\bar{\psi}_j\gamma_k\delta\psi_i+\bar{\psi}_i\gamma_j\delta\psi_k+
\bar{\psi}_j\gamma_i\delta\psi_k)+
$$
$$
+\gamma_n\sigma^{jk}\psi_i(\bar{\psi}_j\gamma_k\delta\psi_l+\bar{\psi}_l
\gamma_j\delta\psi_k+\bar{\psi}_j\gamma_l\delta\psi_k)+\gamma_i\sigma^{jk}
\psi_n(\bar{\psi}_j\gamma_k\delta\psi_l+\bar{\psi}_l\gamma_j\delta\psi_k+
\bar{\psi}_j\gamma_l\delta\psi_k))-
$$
$$
-{i\over{12}}\epsilon^{nml}[\chi_m\Psi]\gamma^i(\gamma_l\psi_s(
\bar{\psi}_n\gamma^s\delta\psi_i)+\gamma_n\psi_s(\bar{\psi}_i\gamma^s
\delta\psi_l)+\gamma_i\psi_s(\bar{\psi}_n\gamma^s\delta\psi_l))\} \eqno(5.12)
$$
From here we can now extract expressions for the electric and spinor 
fields, $e^{ij}\Psi$ and $\chi_i\Psi$, in terms of functional derivatives 
of gauge invariant wave-functionals, with respect to the gauge invariant 
variables. Observe that for that, we have to solve a linear system of 
differential equations, therefore involving the inversion of differential 
operators. We can then conclude that in general, both operators 
$e^{ij}\Psi$ and $\chi_i\Psi$, will depend non-locally on the functional 
derivatives $\delta\Psi/\delta g_{ij}$ and $\delta\Psi/\delta\psi_k$. Like 
in the non-supersymmetric case \cite{kj1}, the Hamiltonian will thus be a 
non-local composite operator.

Before proceeding with the study of these non-local operators, there is one 
more composite operator that we still would like to express in a manifestly 
gauge invariant way, {\it i.e.}, that we would like to geometrize. Such an 
operator is the supersymmetry generator, (3.5-6). In particular, we will 
look at its structure as depicted in equations (3.5),(3.25-26), and geometrize 
the tensor ${\cal q}^i(x)$. For that, we simply have to make use of the 
previous formulae into equation (3.26), and obtain,
$$
{\cal q}^i=\pmatrix{
0 & -{e\over6}\epsilon^{nml}{e^i}_m+{1\over{24eg}}\epsilon^{ijk}({R^{nl}}_{jk}-
{R^{ln}}_{jk}) \cr
{e\over6}\epsilon^{nml}{e^i}_m+{1\over{24eg}}\epsilon^{ijk}({R^{nl}}_{jk}-
{R^{ln}}_{jk}) & 0 \cr}\cdot
$$
$$
\cdot\gamma^r(\gamma_l{\cal D}_r\psi_n+\gamma_n{\cal D}_l\psi_r+
\gamma_r{\cal D}_l\psi_n) \eqno(5.13)
$$
from where the supersymmetry generator then follows, according to 
(3.25) and (3.5).

Some words are now in order, concerning the supersymmetry algebra and its 
quantum field theoretic representation on the geometrized fields. One of 
the elements that is present in ${\cal q}^i$ is the non-local operator 
$e^{ij}$, thus turning the supersymmetry generator into a non-local composite 
operator, when expressed in the geometrical variables. As we will see in 
the following, information about the Green's functions present in this 
operator can be obtained, albeit in a formal way. By this, we mean that an 
explicit construction of these Green's functions can only be obtained 
given a particular geometrical configuration (see \cite{kj2} for this 
same situation in the non-supersymmetric case). Precisely due to this 
reason, it follows that the geometric quantum field theoretic representation 
of the supersymmetry algebra can only be explicitly presented once we have 
chosen a particular geometrical configuration, and so have explicit 
knowledge of the Green's functions present in the non-local operators.

Moreover, the geometric supersymmetry generator includes the Riemann tensor 
which is non-linear in the metric and "gravitino" fields, and their 
derivatives; we would therefore also prefer to have a geometrical 
configuration with a high degree of symmetry (a maximal number of Killing 
vectors), in order to simplify it. An example involving spherical 
geometries, generalizing the one in \cite{kj2} to this supersymmetric 
case, shows how this situation is handled explicitly \cite{moi}.

In the pure Yang-Mills case \cite{kj1}, the calculation of the electric 
field tensor involved the inversion of a differential operator that could 
generically have zero modes. Subtleties associated to the inversion of such 
an operator were later handled with the insertion of a deformation into the 
dreibein defining equation \cite{kj2}. We will now see that in this 
supersymmetric case those problems can be better handled, by computing the 
bosonic Green's function for the electric field tensor $e^{ij}$. We will 
see that we will not need to deform our equations in order to obtain a 
well-defined result. We start by inverting the defining equation for the 
electric tensor (5.6), to obtain,
$$
e^{ij}={1\over{\sqrt{g}}}u^{aj}{\delta\over{\delta A^a_i}} \eqno(5.14)
$$
Recall that through the dreibein defining equation (4.6), the gauge 
connection depends on both the dreibein and the "gravitino" field. 
Therefore, we can further expand the geometric tensor $e^{ij}$ as,
$$
e^{ij}(x)={1\over{\sqrt{g(x)}}}u^{aj}(x)\int d^3y \, ({{\delta u^b_k(y)}
\over{\delta A^a_i(x)}}{\delta\over{\delta u^b_k(y)}}+{{\delta\psi_k(y)}
\over{\delta A^a_i(x)}}{\delta\over{\delta\psi_k(y)}}) \eqno(5.15)
$$
Variations of the dreibein can be further separated into variations of the 
six gauge invariant degrees of freedom $g_{ij}$ and of the three gauge 
degrees of freedom. As we are considering operators that act on gauge 
invariant wave-functionals only, we will simply obtain,
$$
e^{ij}(x)=\int d^3y \, {1\over{\sqrt{g(x)}}} \{ (u^{aj}(x){{\delta u^b_k(y)}
\over{\delta A^a_i(x)}}u^b_m(y))2{\delta\over{\delta g_{km}(y)}}+u^{aj}(x) 
{{\delta\psi_k(y)}\over{\delta A^a_i(x)}}{\delta\over{\delta\psi_k(y)}} \} 
\eqno(5.16)
$$

We now want to study the bosonic Jacobian matrix $\delta u/\delta A$, and 
see that it has a better behavior in here, than in the non-supersymmetric 
Yang-Mills case. For that, we need to start by geometrizing such a matrix. 
Let us re-write (4.6) as,
$$
\epsilon^{ijk}(\delta^{ac}\delta^n_k\partial_j+\delta^n_k\epsilon^{abc}
A^b_j-{i\over4}\delta^{ac}(\bar{\psi}_j\gamma^n\psi_k))u^c_n = 0 \eqno(5.17)
$$
Variation of this equation is (4.8-9),
$$
\epsilon^{ijk}D_j\delta u^a_k=-\epsilon^{ijk}\epsilon^{abc}u^c_k\delta A^b_j+
{i\over2}\epsilon^{ijk}\bar{\psi}_j\gamma^a\delta\psi_k \eqno(5.18)
$$
and to obtain $\delta u^a_i$ in terms of $\delta A^a_i$, the operator 
acting on $\delta u^a_i$ must be inverted. In order to do so, let us 
consider the associated eigenvalue problem,
$$
\epsilon^{ijk}(\delta^{ac}\delta^n_k\partial_j+\delta^n_k\epsilon^{abc}A^b_j-
{i\over4}\delta^{ac}(\bar{\psi}_j\gamma^n\psi_k)){w_A}_n^c=\sqrt{g}
\Lambda_A{w_A}^{ia} \eqno(5.19)
$$
By definition, one solution to this equation with $\Lambda_A=0$ is $u^a_i$ 
itself. In our notation, $A$ labels all the eigenfunctions, except the 
particular one given by $u^a_i$. Moreover, it will be assumed that 
$\{u^a_i,{w_A}^a_i\}$ forms a complete orthonormal spectrum of real 
eigenfunctions for the considered operator. By orthonormality, we mean,
$$
\int \sqrt{g} \, d^3x \, g^{ij} (u^a_i{w_A}^a_j)=0
$$
$$
\int \sqrt{g} \, d^3x \, g^{ij} ({w_A}^a_i{w_B}^a_j)=3V\delta_{AB}
$$
$$
\int \sqrt{g} \, d^3x \, g^{ij} (u^a_iu^a_j)=3V \eqno(5.20)
$$
where $V$ is the volume of the space described by $g_{ij}$ ({\it i.e.}, $V$ 
is a "dynamical" volume), and $\delta_{AB}$ is a Kronecker or Dirac delta, 
depending on wether the spectrum is discrete or continuous. Now, expand a 
generic variation of the dreibein in this complete set,
$$
\delta u^a_i=\eta u^a_i +\sum_A \eta_A {w_A}^a_i \eqno(5.21)
$$
and substitute this in (5.18). If we dot on the left (meaning inner product 
with the required measure (5.20)) with the same complete set 
$\{u^a_i,{w_A}^a_i\}$, we will obtain a non-homogeneous linear system of 
equations for the expansion parameters, $\eta$ and $\eta_A$. Solving that 
system, and inserting the result in (5.21) yields an expansion of the 
variation $\delta u^a_i$ in terms of the variations $\delta A^a_i$ and 
$\delta\psi_k$. It is then easy to compute the Jacobian matrix 
$\delta u/\delta A$. However, such a result will not be naturally geometric, 
as it involves the eigenfunctions ${w_A}^a_i$, which are gauge vectors. To 
solve this problem, we now introduce the geometric modes ${z_{Ai}}^j$, 
associated to the gauge modes ${w_A}^a_i$, and defined via,
$$
{w_A}_i^a\equiv{z_{Ai}}^ju^a_j \eqno(5.22)
$$
It can then be shown that these geometric modes obey,
$$
\epsilon^{ijk}\nabla_j{z_{Ak}}^m=\sqrt{g}\Lambda_A{z_A}^{im} \eqno(5.23)
$$
So, the ${z_A}^{ij}$ are the eigenmodes of the geometric curl operator, with 
the same eigenvalues as the gauge modes ${w_A}^a_i$, $\Lambda_A$. Full 
geometrization of the Green's function,
$$
u^{aj}(x){{\delta u^b_k(y)}\over{\delta A^a_i(x)}}u^b_m(y) \eqno(5.24)
$$
is now at hand. The result is,
$$
u^{aj}(x){{\delta u^b_k(y)}\over{\delta A^a_i(x)}}u^b_m(y) = \sqrt{g(x)}
g^{ij}(x) \Bigl\{ {{\cal H}_{kms}}^s(y,x)-
$$
$$
-{3\over{\int d^3x \, \epsilon^{ijk}(\bar{\psi}_j\gamma_i\psi_k)}}\int 
d^3u \, \epsilon^{ijk}(\bar{\psi}_j\gamma_n\psi_k)(u){{\cal H}_{kmi}}^n(y,u)+
$$
$$
+g_{km}(y) {{8i}\over{\int d^3x \, \epsilon^{ijk}(\bar{\psi}_j\gamma_i
\psi_k)}} \bigl\{3+{i\over8}\int d^3u \, \epsilon^{ijk}(\bar{\psi}_j
\gamma^n\psi_k)(u){{\cal H}_{nis}}^s(u,x) -
$$
$$
-{{3i}\over{8\int d^3x \, \epsilon^{ijk}(\bar{\psi}_j\gamma_i\psi_k)}}\int
\int d^3u \, d^3v \, \epsilon^{ijk}(\bar{\psi}_j\gamma^n\psi_k)(u) 
\epsilon^{mrs}(\bar{\psi}_r\gamma_l\psi_s)(v){{\cal H}_{nim}}^l(u,v) 
\bigr\} \Bigr\} \eqno(5.25)
$$
where we have defined the Green's functions,
$$
{\cal H}_{ijmn}(x,y)\equiv\sum_{AB}{z_A}_{ij}(x)I_{AB}^{-1}{z_B}_{mn}(y) 
\eqno(5.26)
$$
and the matrix,
$$
I_{AB} \equiv -{3\over2}V\Lambda_A\delta_{AB}-{i\over8}\int d^3x \, 
\epsilon^{ijk}(\bar{\psi}_j\gamma^n\psi_k){z_A}_{im}{z_{Bn}}^m+
$$
$$
+{i\over8}{1\over{\int d^3x \, \epsilon^{ijk}(\bar{\psi}_j\gamma_i
\psi_k)}}(\int d^3x \, \epsilon^{ijk}(\bar{\psi}_j\gamma_l\psi_k)
{z_{Ai}}^l)(\int d^3x \, \epsilon^{ijk}(\bar{\psi}_j\gamma^n\psi_k)
{z_B}_{ni}) \eqno(5.27)
$$

We see that we have obtained a well-defined result, unlike it would have 
happen in the non-supersymmetric case \cite{kj1}, where there were 
divergences in the electric energy that were independent of the geometry. 
Clearly, the Green's functions (5.26) may still have geometry dependent 
divergences associated with the degree of symmetry of a given geometrical 
configuration (as determined by its Killing vectors). Also, this may seem a 
somewhat formal result, but observe that now we have a constructive 
definition of the Green's function (5.24): given a geometrical configuration 
on the domain manifold, we start by solving the eigenvalue equation in order 
to obtain the geometric eigenmodes. Once we have such eigenmodes, we first 
construct the matrix $I_{AB}$, then invert it (most likely through a 
symbolic manipulation program), and finally compute the Green's functions 
${\cal H}_{ijmn}(x,y)$.

We have now geometrized all the tensors appearing in the Hamiltonian 
formulation of $N=1$ supersymmetric Yang-Mills theory. All composite 
operators should now follow in a straightforward fashion.

\sect{Conclusions}

We have defined new local gauge invariant variables for supersymmetric 
gauge theory. These variables have moreover a geometrical interpretation, 
as they are a three dimensional "metric" and a "gravitino". The geometry 
associated to the 
theory is then just like the geometry of supergravity.

We have also shown that these new variables are free of Wu-Yang ambiguities; 
so they seem to be quite appropriate for the study of nonperturbative 
phenomena in supersymmetric gauge theories, as they explicitly parameterize 
the physical Hilbert space of the theory. We have also seen that these 
variables have a better behavior here than in the non-supersymmetric case. 
Namely, there are no geometry independent divergences in the bosonic half 
of the electric energy tensor operator.

The treatment presented here was rather formal, and the issue of 
renormalization was not addressed. Further work on this formalism should 
focus on this problem. We could think of using the known beta functions of 
supersymmetric Yang-Mills theory and perform the renormalization of the 
(geometric) composite operators that we have presented. These renormalized 
operators could then be used to extract information on the ground state 
wave-functional of the gauge theory. In this supersymmetric case, the 
functional differential equation for the ground state is $Q\Psi[g,\psi]=0$, 
which is first-order in the non-local functional derivatives. There would 
seem to be hope that we could then extract some information about the 
solution to the theory.

It would also be interesting to study special solutions to supersymmetric 
Yang-Mills theory in this framework. Namely, we could try to extend to this 
supersymmetric case the example of spherical geometries that was introduced 
in \cite{kj2}. In particular, in order to define a vector-spinor on a 
three manifold, the manifold must be parallelisable, and its second 
Stiefel-Whitney cohomology class must be trivial. Such is the case for 
$S^3$, so that the example in \cite{kj2} could indeed be generalizable to 
this framework \cite{moi}. There is also the possibility of extending this 
formalism to 
higher $N$ supersymmetric gauge theory. This could be interesting, 
specially if some connection to the work in \cite{sw} could then be 
established. 

All these lines of work are quite interesting to follow as there is good 
knowledge about some properties of supersymmetric Yang-Mills theories (see 
\cite{shif} for a modern review, and references therein). In the example 
of spherical geometries, a bridge between our formalism and the well 
known instanton solutions of gauge theory can be established; while in 
the case of extended $N=2$ supersymmetric Yang-Mills theory, we can observe 
the interesting fact that BPS states obey -- in the geometrical formulation 
-- a three dimensional "Einstein field equation" where the "stress tensor" 
is the one associated to a gauge vector field \cite{moi}.

One well known -- and exact -- result from supersymmetric Yang-Mills 
theory is that the gluino condensate $\langle\lambda\lambda\rangle$ develops 
a non-zero vacuum expectation value, thus spontaneously breaking the 
discrete symmetry of the model, $Z_{2N}$, down to $Z_2$. There will then be 
$N$ degenerate vacua, all with vanishing energy density \cite{shif}. In 
order to reproduce such a result in this geometrical formalism we need 
to set up the calculation of the gluino condensate, and for that we need 
a trial wave-functional in order to obtain the $\langle\lambda\lambda\rangle$ 
vacuum expectation value in this Hamiltonian formulation. So again, we will 
have to restrict ourselves to a particular geometrical configuration 
in order to perform the calculation. An interesting fact here is that 
the result for the gluino condensate \cite{shif} comes from a one-instanton 
calculation, and so it would be very interesting to know if the spherical 
geometry configuration can reproduce such a result. We hope to report on 
these questions in the near future \cite{moi}.

\renewcommand{\thesection}{}
\sect{\hspace{-1em}Acknowledgements}
I would like to thank Kenneth Johnson for many discussions and advice, and 
Peter Haagensen for comments and reading of the manuscript. The author is 
partially supported by the Praxis XXI grant BD-3372/94 (Portugal).

\end{document}